# Hydrogen nanobubbles in a water solution of dietary supplement

**Vladimir L. Safonov** [a,b] **and Anatoly K. Khitrin** [c]

**Using gas chromatography, proton nuclear magnetic resonance and qualitative experiments, we demonstrate that a water solution of dissolved dietary supplement, creating negative redox potential, contains invisible hydrogen nano-bubbles, which remain in the solution for several hours.**

Hydrogen, as it was recently revealed by basic and clinical research, is an important physiological regulatory factor producing antioxidant, anti-inflammatory and anti-apoptotic protective effects on cells and organs.[1] Drinking hydrogen-rich water is shown to be one of safe practical methods of hydrogen therapy. However, the solubility of hydrogen in water is very low, about $1.6 \cdot 10^{-4}$ g (or, 1.84 mL of gas at 1 atm) in 100 g of $H_2O$ at 20° C. In order to increase the amount of hydrogen in water, one can try to add more hydrogen gas in a form of small bubbles. Nanobubbles can be stabilized by a balance between surface tension and repulsive forces between surface electric charges, as it has been discussed before.[2] In this Letter we demonstrate that hydrogen nanobubbles can exist in the water solution of a dietary supplement for a sufficiently long time.

The dietary supplement was developed as a chemical composition for producing stable negative oxidation-reduction potential in drinkable liquids.[3] Our samples were prepared by dissolving effervescent tablets containing potassium bicarbonate, sodium bicarbonate, magnesium particles (~40 μm), tartaric acid, L-leucine, organic sea salt, calcium lactate and inulin. One 230 mg tablet, fully dissolved within 5 minutes in a 0.5 L of purified water, provides a minimum of 8.5 pH alkalinity and about -500 mV of redox potential. Reactions of magnesium particles with chemicals of the tablet and water produce hydrogen gas.

In the first one - two minutes of reaction, dissolution of the tablet in water occurs quite rapidly and is accompanied by a large number of bubbles. Then, the bubbling process visually stops after about 5 minutes, demonstrating a complete dissolution of the tablet in water. The water solution becomes completely transparent with random bubbles on the glass walls. No specific activity was observed after that except for rare small bubbles appearing from time to time in the volume and then going out of the solution.

[a] *Mag and Bio Dynamics, Inc., Escondido, California 92029, U.S.A.;*
*E-mail: vlsafonov@magbiodyn.com*

[b] *Center for Personalized NanoMedicine at Florida International University, Miami, Florida 33199, U.S.A.*

[c] *Department of Chemistry, Kent State University, Kent, Ohio 44242, U.S.A.; E-mail: akhitrin@kent.edu*

*No financial disclosures were reported by the authors of this paper.*

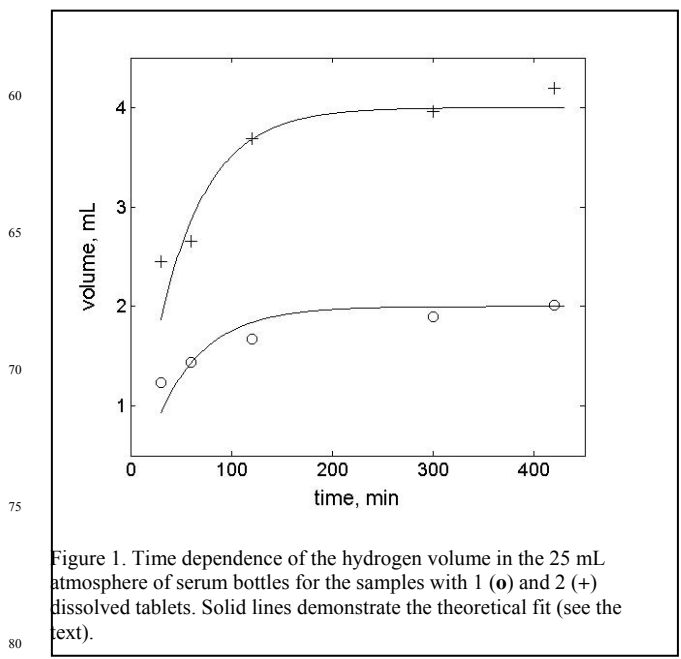

Figure 1. Time dependence of the hydrogen volume in the 25 mL atmosphere of serum bottles for the samples with 1 (**o**) and 2 (**+**) dissolved tablets. Solid lines demonstrate the theoretical fit (see the text).

First, we performed a qualitative experiment, which suggests that our solution may contain invisible bubbles. Few drops of olive oil have been added to the solution. Then the glass was shaken to mix the oil with the solution. The result we observed was oil emulsification, which is not observed if pure water is used. One of possible explanations is that emulsification is facilitated by a presence of interface (gas-liquid) boundaries in a bulk of our system.

Pure solution remains transparent, without producing noticeable light scattering. Therefore, the bubbles, if they exist, should have sizes smaller than the wavelength of visible light λ=380 - 780 nm.

In order to have a more direct experimental proof of our assumption, we have measured the hydrogen gas ($H_2$) yield from the tablets dissolved in water at room temperature. The experiment was conducted in 125 ml serum bottles. Three samples were prepared by dissolving 0, 1, or 2 tablets. The tablets were dropped and dissolved in 100 ml of double-distilled water. The bottles were closed with butyl rubber stopper and sealed with aluminium seals. The rest of bottle volume of 25 ml was analysed for the presence of hydrogen gas. The analysis was performed by taking 1 mL of head space atmosphere by 1 mL syringe. Then this 1 mL gas sample was injected into 10 mL analytical vial to analyze hydrogen percentage using gas chromatograph (Agilent, micro GC 3000). The fluctuation of hydrogen content in the control sample without dissolved tablet was not greater than 0.044%.

The time dependence of hydrogen volume in the atmosphere of serum bottles for the samples with one and two dissolved tablets is shown in Figure 1. We see that it takes several hours before the hydrogen yield slowly comes to a stop. The time dependences of hydrogen yield volume from the solution with one tablet and from the solution with two dissolved tablets look similar, except that the hydrogen yield for the two-tablet solution is nearly doubled. We have approximated the volume time dependence as

$$V(t) = V\left[1 - \exp\left(\frac{t}{\tau}\right)\right] \qquad (1)$$

A fit for the lower curve was obtained with $V$ = 2 mL and $\tau$ = 47.6 min. For the upper curve the maximum volume was doubled ($V$ = 4 mL) and the same characteristic time $\tau$ = 47.6 min was used. A single tablet produces more hydrogen than needed to reach equilibrium saturation described by the Henry's law. Therefore, for equilibrium saturation, one would not expect significant difference between dissolving one or two tablets. The experiment in Figure 1 shows that the dissolved hydrogen is present in over-equilibrium concentration. Again, a plausible explanation is that it may form nanobubbles.

Let us consider a simple quantitative model which allows estimation of the lifetime of a nanobubble in water solution. We can use the equality of the Archimedes force (which is proportional to the bubble volume) and the Stokes' drag, the frictional force acting on a moving spherical bubble:

$$\tfrac{4}{3}\pi r^3 \rho g = 6\pi r \eta v. \qquad (2)$$

Here $r$ is the bubble radius, $\rho$ is the water density, $g$ is the free fall acceleration, $\eta$ is the dynamic viscosity and $v$ is the bubble velocity. From equation (2) one can easily find the radius as

$$r = 3\sqrt{\eta v / 2\rho g}\ . \qquad (3)$$

Taking $\eta$ = 1.0 $10^{-3}$ Ns/m$^2$, $\rho$ = $10^3$ kg/m$^3$, $g$ = 9.8 m/s$^2$ and $v$ = 1 cm/hour, we obtain $r$ = 1.1 μm. Smaller bubbles will stay in the bulk for hours before reaching the upper surface. In this case, the major mechanism of their motion will be a convective flow of the liquid.

To provide further evidence for the presence of hydrogen nanobubbles we have performed nuclear magnetic resonance (NMR) study. The idea of NMR experiment is the following. Water is a diamagnetic substance with magnetic susceptibility (volumetric susceptibility) $\chi$ = -9.0 ppm in SI system. It is a dimensionless quantity related to the relative permeability $\mu$ as $\chi = \mu - 1$. Because of different magnetic susceptibilities of water and hydrogen gas (the latter is negligible), the bubbles would create inhomogeneous magnetic field in the surrounding liquid. This inhomogeneity of magnetic field can be measured as broadening of NMR line. Distribution of magnetic fields and, therefore, NMR line shape can be calculated by using Anderson's statistical theory (its details can be found in Abragam's book [4]). It follows that the relative broadening of NMR line $\Delta f / f$ is proportional to the relative volume $\Delta V / V$ occupied by the bubbles:

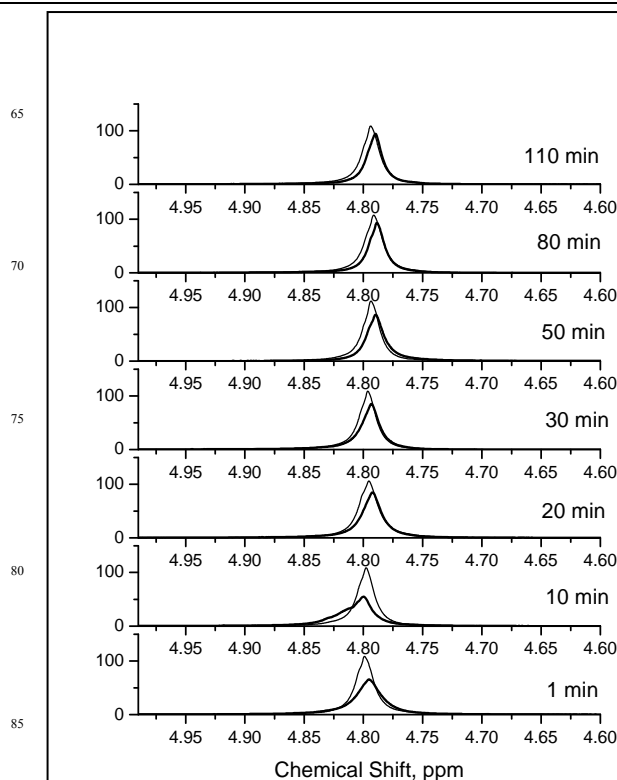

Figure 2. NMR proton spectra (bold lines) at different times after adding the tablet material to the solvent. Thin lines are the spectra of the control sample.

$$\Delta f / f = 1.2\,|\chi|\,(\Delta V / V). \qquad (4)$$

The sample was 4 mg of the tablet material dissolved in 1 mL of the mixture 10% $H_2O$ and 90% $D_2O$ (heavy water) placed in 5 mm NMR tube. The control sample was 1 mL of the same mixture in a separate 5 mm NMR tube. Proton ($^1H$) NMR spectra have been recorded using 500 MHz Varian Unity/Inova NMR spectrometer at 23° C. Spectra at different times after adding the tablet material to the solvent (bold lines), together with the spectra of the control sample (thin lines), recorded 1 min before, are shown in Figure 2.

We see that compared to the control, the spectrum after 1 min is significantly broadened. Even stronger broadening is observed after 10 min. Macroscopic bubbles, when present on the walls of the NMR tube, show themselves on the right side of the spectrum as irregular peaks with up to 0.1 ppm diamagnetic shifts. Bubbles on the walls have been removed each time immediately before the spectrum acquisition. After 20 min and longer, all spectra show about 0.004 ppm diamagnetic shift compared to the control sample. After 24 hours a set of four pairs of measurements with 1 min interval between acquisitions has been performed. It showed a consistent diamagnetic shift by 0.004 ppm of the sample with dissolved tablet material compared to the control sample. The peak positions for all eight spectra were reproducible within 0.001 ppm. Small diamagnetic shift can be attributed to the effect of dissolved tablet material and to elimination of residual oxygen.

After 20 min and beyond, the additional line broadening decays exponentially, as one can see in Figure 3. We used the following fitting dependence:

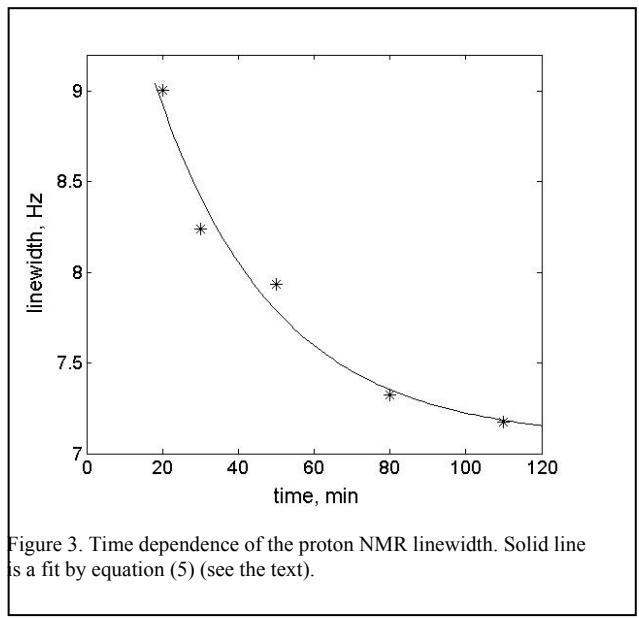

Figure 3. Time dependence of the proton NMR linewidth. Solid line is a fit by equation (5) (see the text).

$$\Delta f(t) = \Delta f_0 + \Delta f_1 \exp\left(-\frac{t}{\tau_f}\right), \quad (5)$$

where $\Delta f_0$ is the proton NMR linewidth of the water solution without nanobubbles, $\Delta f_1$ is the additional broadening, created by nanobubbles and $\tau_f$ is the characteristic decay time. The best fit with $\Delta f_0$ = 7.08 Hz, $\Delta f_1$ = 3.48 Hz and $\tau_f$ = 31.4 min is shown by solid line in Figure 3. We can notice that the decay time for NMR line broadening, which, according to equation (4), can be interpreted as the characteristic time of the bubbles volume decrease, is close to the characteristic time $\tau$ = 47.6 min obtained for the hydrogen yield (Figure 1).

By using equation (4), we can also estimate the relative volume of the bubbles. For the spectrum after 10 min the result is $\Delta V / V \approx 10^{-3}$. This ratio is one order of magnitude less than the hydrogen yield in Figure 1. This discrepancy may come from two reasons. First, the actual volume $\Delta V$ can be greater than our estimate because a self-diffusion of water molecules can partially average inhomogeneity of magnetic field and reduce the line broadening. Estimation of this effect requires knowledge of actual size of nanobubbles. Second, the hydrogen pressure in nanobubbles can be greater than the atmospheric pressure [5-7] and, therefore, the atmospheric-pressure volume of hydrogen in nanobubbles is greater than the volume of bubbles.

We can mention that our results for long-lived hydrogen nanonbubbles are consistent with the characteristics of hydrogen nanobubbles in solutions obtained by water electrolysis [6]. The method, using NMR spin-lattice relaxation, has been previously used to show the presence of stable air and Xe nanobubbles in water.[7] Our approach, summarized by equations (4) and (5), provides more details: one can estimate both the effective volume of nanobubbles and the time evolution of this volume.

There is a possibility that very small (few nanometers) hydrogen nanobubbles exist even after 24 hours in the water solution, which still has a negative redox potential. Complex microscopic structure of water[8] should greatly affect the existence and the properties of such small nanobubbles.

## Conclusions

In summary, we provided experimental evidence that a dietary supplement, designed to saturate water with hydrogen, produces long-lived hydrogen nanobubbles. Proton nuclear magnetic resonance is shown to be a useful method of estimating both the volume of nanobubbles and their lifetime in the solution.

## Acknowledgements

The Authors wish to thank Prof. Dusan Miljkovic for helpful discussions and Dr. Anna Obraztsova for technical help. Samples were provided by DIBAL LLC (San Diego, California). Material of samples was a dietary supplement in the form of effervescent tablets AquaActive with an elemental magnesium blend. Currently the material is also available as an oral-ingestible dietary supplement called Recovery with HydroFX.